\documentstyle[12pt]{article}

\everyjob{\typeout{MuTeX.sty: LaTeX Modification by EPM.}}
\immediate\write10{MuTeX.sty: LaTeX Modification by EPM.}

\makeatletter   

\input{ifthen.sty}

\newtheorem {theorem} {Theorem} [section]

\newtheorem {Canham threshold} [theorem] {Canham Threshold}

\def\theoremstyle#1#2{\def\@@theoremheadstyle{#1}
                      \def\@@theorembodystyle{#2}}
\def\@@theoremheadstyle{\sc}
\def\@@theorembodystyle{\rm}
\def\@begintheorem#1#2{\@@theorembodystyle 
                       \trivlist 
		       \item[\hskip 
                             \labelsep{\@@theoremheadstyle #1\ #2}]}
\def\@opargbegintheorem#1#2#3{\@@theorembodystyle 
                              \trivlist 
			       \item[\hskip 
				  \labelsep{\@@theoremheadstyle #1\ #2\ (#3)}]}

 \def\@@pc{\bf}

 \newcommand {\pcodestyle}[1] {\def\@@pc{#1}}  

 {
 \def\PROGRAM		{{\@@pc program\ }}
 
 \def\PROCEDURE		{{\@@pc procedure\ }}
 \def\FUNCTION		{{\@@pc function\ }}
 \def\LOCAL		{{\@@pc local\ }}
 \def\GLOBAL		{{\@@pc global\ }}
 \def\RETURNS		{{\@@pc returns\ }}
 \def\RETURN		{{\@@pc return\ }}
 \def\BEGIN		{{\@@pc begin\ }}
 \def\END		{{\@@pc end\ }}
 \def\IF			{{\@@pc if\ }}
 \def\THEN		{{\@@pc then\ }}
 \def\ELSE		{{\@@pc else\ }}
 \def\REPEAT		{{\@@pc repeat\ }}
 \def\UNTIL		{{\@@pc until\ }}
 \def\WHILE		{{\@@pc while\ }}
 \def\DO			{{\@@pc do\ }}
 \def\FOR		{{\@@pc for\ }}
 \def\TO			{{\@@pc to\ }}
 \def\DOWN		{{\@@pc down\ }}
 \def\NEXT		{{\@@pc next\ }}
 \begin{center}
  \begin{minipage}{.9\textwidth}
   \small
    \begin{tabbing}
 }%
 {  \end{tabbing}%
   \end{minipage}%
  \end{center}%
 }

			   {\end{quotation}}

			 {  \end{minipage}%
			   \end{center}%
			  \end{description}%
			 }

				{ \end{minipage}
				 \end{center}%
				}

\def\thebibliography#1{\section*{References}\list
 {[\arabic{enumi}]}{\settowidth\labelwidth{[#1]}\leftmargin\labelwidth
 \advance\leftmargin\labelsep
 \usecounter{enumi}}
 \def\newblock{\hskip .11em plus .33em minus -.07em}
 \sloppy
 \sfcode`\.=1000\relax}

\newsavebox{\ProofSym}
\savebox{\ProofSym}{%
  \begin{picture}(10,10)
    \put(0,0){\framebox(9,9){}}
    \put(0,3){\framebox(6,6){}}
  \end{picture}}
\newcommand{\eop}{\hfill\usebox{\ProofSym}}
\newenvironment{proof}{\noindent {\sc Proof.\/}}{\eop\par\vspace{0.3cm}}

\begin{document}

\title{Multi-partite Quantum Entanglement versus 
Randomization: Fair and Unbiased Leader Election in Networks}

\author{
Sudebkumar Prasant Pal\\email:spp@cse.iitkgp.ernet.in\\
http://www.angelfire.com/or/sudebkumar\\
Department of Computer Science and Engineering\\
Indian Institute of Technology, Kharagpur- 721302,
India \\
\\Sudhir Kumar Singh\\email:sudhirks@ieee.org, sudhirks@computer.org\\
\and Somesh Kumar\\email: smsh@maths.iitkgp.ernet.in\\
\and Department of Mathematics\\
Indian Institute of Technology, Kharagpur- 721302, India\\
}

\date{}

\maketitle


\begin{abstract}

In this paper we show that sufficient multi-partite quantum entanglement
helps in fair and unbiased election of a leader 
in a distributed network of
processors with only linear classical communication complexity. 
We show that a total of $O(\log n)$ distinct
multi-partite maximally entanglement sets (ebits)
are capable of supporting such a protocol 
in the presence of nodes that may lie and thus be biased.
Here, $n$ is the number of nodes in the network.
We also demonstrate the difficulty of performing unbiased and fair election of
a leader with linear classical communication complexity 
in the absence of quantum entanglement even if all nodes have 
perfect random bit generators. We show that the presence of a sufficient 
number $O(n/\log n)$ of biased agents leads to a non-zero 
limiting probability of biased
election of the leader, whereas, the presence 
of a smaller number $O(\log n)$
of biased agents matters little. We 
define two new related complexity classes
motivated by the our leader election problem 
and discuss a few open questions.

\end{abstract}

    \textwidth 6.6in
    \textheight 8.5in

\section{Introduction}
\label{introduction}

{\it Quantum entanglement} is one of the most fascinating 
aspects of quantum physics. Two 
particles in an entangled state behave as if possesing a common 
state, even if they are physically separated over a great
distance. Two maximally entangled qubits can be in the Bell 
state $|\Phi ^+>={\frac 1 {\sqrt 2}} (|00>+|11>) $. Each of the qubits 
here is in a mixed state with probability ${\frac 1 2}$ in state $|0>$ and
with probability 
${\frac 1 2}$ in state $|1>$. However, both together form a pure state 
\cite{gruska}. Such a pair of qubits is called an EPR pair (after
Einstein, Podolsky and Rosen) \cite{EPR1935,gruska}.
Such entangled qubits can exhibit what physicists call {\it nonlocal effects}.
Nonlocal effects cannot
occur between particles as per laws of classical physics unless communication
is permitted between the physically separated particles, 
in which case communication would have to be at a speed 
exceeding that of light.
Such behaviour was referred to as {\it spooky actions at a 
distance} in \cite{EINSTEIN1971} and was earlier 
mentioned in the celebrated 1935
paper by Einstein, Podolsky and Rosen \cite{EPR1935}. 
Nonlocal effects however do not involve any communication and therefore
the question of faster than light communication does not arise. Even if 
two agents Alice and Bob share entangled pairs, there is no way for Alice to 
manipulate her own particles to convey any information, she has, to Bob; 
information must be sent explicitly from Alice to Bob.
In this context, it is necessary to investigate whether quantum computing 
resources help Alice and Bob solve a problem together with lesser communication
complexity than that required when no quantum resource is used.

In Yao's model \cite{yaoqc}, Alice and Bob are allowed to communicate
using qubits. Kremer's result in \cite{kremer} does not show superiority
of quantum communication over classical communication for the 
INNER PRODUCT function $f(x,y)=x.y$ where $x$ and $y$ are $n$-bit vectors
each, $x$ given only to Alice as input and $y$ given only to Bob as input; 
the lower bound on qubits required for computing 
INNER PRODUCT is $\Omega (n)$. 
However, the result of Buhrman, Cleve and Wigderson 
\cite{BCW1998} requires only
$O(\sqrt n \log n)$ qubit communication cost for the DISJOINT 
function, whose classical probabilistic communication complexity is 
$\Omega (n)$ \cite{KS1992,R1992}. The DISJOINT(x,y) function gives 0 
if any bit of $n$-bit vector $x$ is equal to the
corresponding bit of the $n$-bit vector $y$; otherwise, it returns a 1.

Just as two particles at a distance could be quantum entangled forming
an EPR pair, 
it is also possible to entangle multiple mutually physically 
separated particles. 
One example is due to
Greenberger, Horne and Zeilinger \cite{GHZ1989,intquantumeditedbook}; 
here, three particles are quantum entangled. 
Singh, Kumar and Pal \cite{SKP2003} develop two protocols that 
entangle three nodes $A,B$ and $C$ in a network
in the GHZ state ~~${({|0_A0_B0_C>+|1_A1_B1_C> })/  {\sqrt 2}}$~~ 
if two pairs (say $(A,B)$ and $(B,C)$)
were earlier entangled as two EPR pairs. The first protocol uses the standard
teleportation circuit \cite{nc} and the second one uses a new circuit
developed in \cite{SKP2003}. Both protocols use only classical communication.
Buhrman, Cleve and van Dam \cite{BCD2001},
make use of three-party entanglement and demonstrate the existence of a
function whose computation requires strictly lesser classical 
communication complexity 
compared to the scenario where no quantum entanglement is used.

Singh, Kumar and Pal \cite{SKP2003} show how $n$ nodes connected in a network
can be entangled together in a maximally entangled $n$-partite state 
${({|0_10_2...0_n>+|1_11_2...1_n>}) {\sqrt 2} }$
using 
linear classical communication complexity provided the nodes were earlier 
entangled in $(n-1)$ EPR pairs. The $(n-1)$ EPR pairs given initially are
required to be shared between agents forming edges of a spanning tree in 
the network; such a spanning tree is called the {\it 
spanning EPR tree} \cite{SKP2003}.
Brassard et al. show in \cite{BBT2003} that prior multi-partite entanglement 
can be used by $n$ agents in a {\it perfect} quantum protocol (not using 
any form of communcation between the $n$ agents),
to solve a multi-party
distributed problem, whereas, no classical deterministic protocol succeeds in
solving the same problem with probability away from $\frac 1 2$ by a fraction
that is larger than an inverse exponential in the number of agents.
Multi-partite entanglement is also used by Buhrman, van Dam, Hoyer and Tapp
\cite{BDHT1999} where an $n$-party problem was shown to have quantum
communication complexity $n$ qubits, 
whereas the classical communication complexity was
shown to be $\Omega (n\log n)$ cbits. 


In this paper we show that a sufficient amount of multi-partite 
quantum entanglement
helps in electing a leader unambiguously in a distributed network of agents
where it is not posible for any agent to introduce any statistical 
{\it bias} in the election.
We show in our Protocol I that a total of $O(\log n)$ distinct
multi-partite entanglement states are capable of supporting a
protocol with linear classical communication complexity
for leader election even in the presence of nodes that may lie or be 
{\it biased}.
Here, $n$ is the number of nodes in the network. Unbiased election is 
enabled by the sharing of multiple multi-partite entangled states. Using one
honest referee, we also ensure that all honest nodes agree on the 
elected leader.
We also demonstrate the difficulty of performing unbiased and fair election of
a leader with linear classical communication complexity
in the absence of quantum entanglement, even if all nodes have 
perfect random bit generators. We demonstrate this through our Protocol II
where we show the following separation result: 
too many biased liars would prevent the referee from 
conducting a fair and unbiased election, whereas, a small number of liars 
would matter little. This separation result has two aspects. Firstly, it 
shows as expected that the probability of the
election being biased is small and indeed 
vanishes as the number of agents grows asymptotically when
there is a small number of biased agents. So, we observe that it is 
possible to conduct
fair and unbiased election in our randomized Protocol II with 
small error probability
with linear communication cost. Protocol I performs fair and unbiased
election but needs
$O(\log n)$ multi-partite maximally entangled states or $ebits$. 
Note however that even though 
the number of biased voters ($\log n$) is small,
it is an increasing function of $n$. So, with small but growing number of 
liars, the randomized Protocol II achieves vanishingly small 
biased election probability as $n$ grows asymptotically; this demonstrates 
partial tractability of our leader election problem with pure 
randomization and no quantum resource. 
Secondly, the finite limiting probability $e^{-C}, C>0$ ($C$ is a constant), 
of fair and unbaised election with the number 
of biased voters as high as $n/\log n$ (Protocol II)
shows that randomization cannot handle too many cheating agents who are 
essentially biased in a statistical sense. 
Even here, we keep the large number of biased agents $o(n)$ by choosing 
$n/\log n$ biased agents. 

Further, we define two complexity classes (i) $QECC(g,n)$, the class of
problems solvable using $g$ $ebits$ shared amongst $n$ agents with $O(n)$
cbit communication complexity, and, (ii) $RCC(r,n)$, the class of 
problems solvable using $r$ random bits shared bewteen $n$ agents, 
with $O(n)$ cbit communication complexity. We show that $RCC(\log n,n)\subseteq 
QECC(\log n, n)$. The problem of determining whether this is a proper 
containment is an 
interesting fundamental open problem concerning the 
power of quantum entanglement as a computational and communication 
resource.

Throughout our analysis, we follow a uniform 
statistical notion of {\it fairness} and {\it bias}, 
as we define shortly below. In short, an unbiased agent is supposed to 
always vote a 0 or a 1 with equal probability as per an ideal 
random bit generator that the agent possesses. So, any biased agent, using
differing probabililties for 0 and 1 votes, would cause some bias towards some
agent addresses in the election of the leader. Fairness is ensured by giving 
agents equal opportunity in voting for determination of the leader.  

On the 
one hand, the successful completion of a protocol would require all agents
in the network to {\it agree} 
on an elected leader. On the other hand, agents may 
be biased in voting for a leader. 
It is therefore also a matter of concern that
one agent amongst all in the network must act as a {\it referee}, trying to 
conduct a fair and unbiased election process. 
Since election essentially requires voting by all agents in a fair 
setup and, agreement by all agents about the elected leader, we observe that 
linear classical communication complexity is essentially a lower bound in our 
model. In this sense, our model differs from the models of 
Yao \cite{yaocl,yaoqc} and 
also those of \cite{BCD2001,BDHT1999}.

In Section \ref{framework} 
we introduce our own definitions and notions of 
{\it fairness}, {\it bias}, {\it referee} and {\it agreement}. 
We essentially propose two leader election protocols, both use 
only linear amounts of classical communication. 
The first one (Protocol I in Section \ref{quantumelection})
assumes the 
presence of a sufficient number of $n$-partite maximally entangled states, 
and, the second one (Protocol II in Section \ref{randomizedelection})
assumes no such entanglement but only perfect random 
bit generation capability in each agent. We show a separation result 
by demonstrating that too many biased liars would prevent the referee from 
conducting a fair and unbiased election, whereas, a small number of liars 
would matter little. In Section \ref{newclass} we define two new complexity 
classes related to the problem of 
unbiased election of a leader and pose a few open problems.

\section{The Communication Framework}
\label{framework}

We wish to have a complete protocol with a 
starting signal going to each agent, signifying initiation.
This requires communicating a constant number of bits 
of classical information ({\it cbits}) from one special agent to the
rest. (There may be a constant number 
of special tasks like the starting and ending of protocols. So a constant 
number of bits are sufficient for such signals.) We fix one 
arbitrary node to initiate the protocol by sending
this initial message to all other agents. This node is also set as 
the {\it referee} node; the referee is supposed to be conducting the protocol
and collecting {\it agreements} from all agents, including itself, about the 
elected leader. If an agent does not accept the elected leader, he is simply
left out; further computation can be carried out by only those agents who  
accept the election of the leader. In other words, we require total 
agreement over
the elected leader by all the agents if all of them wish to continue 
computations together, after leader election. Such an agreement comes in the 
form of a cbit from each agent after the leader is elected. The referee first 
informs the leader about his election 
and the leader then sends a constant number of cbits to 
each agent, seeking agreement or approval.   

Identifying a leader amounts to fixing 
$\lceil \log n\rceil$\footnote{All logarithms in this paper are
to the base 2.} bits to 
uniquely address an agent out of $n$ agents. 
Any protocol determining a leader must therefore 
come up with these bits. We assume without loss of generality that $n$ is a
power of 2 and therefore we need to determine exactly $\log n$ bits.

Now we state the communication and message passing model for classical 
communication in our protocols. We assume that units of information 
(cbits) are communicated from a source agent $S$ to a destination agent $T$.
The destination can store as many addresses of senders of incoming 
messages so that in future it may choose to reply to any of them. We also 
assume that the cost of communicating a single bit of information 
between any source $S$ and any 
destination $T$ is the same and is counted as one cbit, independent of the
network connecting the agents. 
We are now in a position to derive our protocols.

We define {\it bias} as follows. If an agent $A_i$ can somehow influence the 
election of the leader in a way that is {\it statistically different} 
from the way another  
agent $A_j$ influences election, then we say that 
there is a bias in the election process. In other
words, the determination of the address of the leader must be equally
influenced by each participating agent.

The notion of {\it fairness} is subtlely distinct from that of 
bias. Here, {\it fairness} means ensuring that all agents get a fair
deal in participating in the election process, in a statistical sense.
These notions of statistical bias and fairness will become more explicit 
when we analyse our protocols in the next two sections.

\section{Fair and Unbiased Leader Election Using $n$-partite Entanglement}
\label{quantumelection}

We assume that the $n$ agents $A_1,A_2,...,A_n$
are in multiple $n$-partite maximally entangled 
states presented in the computational bases. Such states are called 
$ebits$ \cite{gruska}. 
Each agent houses his own qubits in these entangled states. 
We also assume that there is a sufficiently large number of such states. 
In particular, we choose to have $\log n$ such 
$ebit$ states $S_1, ..., S_{\log n}$, with each agent $A_i,1\leq i\leq n$ 
housing 
$\log n$ qubits $qu_{i,j},1\leq j\leq \log n$, 
one for each of the $\log n$ states. The state $S_k$ is therefore
${( {|0_{1,k}0_{2,k}...0_{n,k}>+|1_{1,k}1_{2,k}...1_{n,k}>}) /{\sqrt 2} }$ 
for $1\leq k\leq \log n$, where the subscript $k,i$ denotes the $k$-th qubit
of the $i$-th agent. Clearly, a total of $n\log n$ qubits are used.
It is possible to create such a maximally entangled 
multi-partite (ebit) state
using a protocol in \cite{SKP2003} that requires linear classical 
communication complexity and uses up $(n-1)$ EPR pairs generated apriori 
in the form of an {\it EPR spanning tree} of the network. 
Our Protocol I is initiated 
by the referee who sends one cbit to all agents signalling the start of the 
protocol. On getting this trigger, each node $A_i,1\leq i\leq n$ 
performs measurements
on its $\log n$ qubits $qu_{i,j},1\leq j\leq \log n$. These measurements
are performed in the respective {\it computational bases} 
for each of the $\log n$ entanglement sets.
Whoever measures
a qubit, gets a value zero or one, with probability ${\frac 1 2}$ 
for each of 0
and 1 values. Indeed, this is a
perfect random bit generator. Also, no matter in what order $A_i$ 
and $A_j$, $1\leq j<i\leq n$ measure their $k$-th qubits, both of them 
get the same answer. This follows from the postulates of quantum mechanics and 
the property of maximal entanglement for state $S_k$.
The referee can now use these $\log n$ measured bits to produce an address
or index $m$
for the agent $A_m,1\leq m\leq n$
and accept it as an {\it unbiased} leader. Note that such a leader is
perfectly random and therefore totally unbiased. There is
no hiding of the measurements made by the agents about their $k$-th qubits 
since the referee can always check the measured value of the $k$-th qubit of 
the $k$-th entangled state $S_k$ by reading its own $k$-th qubit. Even 
if the referee seeks this value from another agent and the agent lies, 
the referee can simply ignore any false information provided
because it knows the true value of the measured $k$-th qubit. Now 
it is the turn of all agents to send one cbit to the referee and express 
their agreement over the elected leader. If an agent agrees on the leader by 
sending a cbit 1, it is fine and the agent henceforth sticks to the commitment
of having agreed on the leader. Otherwise, we say that the agent is discarded 
and only agreeing agents are considered to contuinue meaningful computations 
any further. We also assume that the elected leader is committed to act
as leader henceforth.
We summarize the lack of bias and the guarantee of fairness in 
the election of the leader in the above Protocol I in the following theorem.

\begin{theorem}
It is possible to conduct an unbiased and fair election of a leader amongst $n$
agents who share $\log n$ maximally entangled $n$-partite states 
with $O(n)$ bits of classical communication.
\label{quantumprotocoltheorem}
\end{theorem}

\section{Randomized Election}
\label{randomizedelection}
We believe that randomization alone is not capable of supporting
distributed protocols that 
can achieve fair and unbiased leader election
provided only a linear number of cbits are exchanged in the protocol. 
Although multi-partite entanglement helps us
perform fair and unbaised leader election with 
only linear classical communication complexity 
(as we see in Protocol I in the previous section), 
randomization alone, coupled with
only linear number of cbits, may lead to fair and unbiased 
election of a leader only in certain special and restricted 
cases. Recall that the leader selected in the case of Protocol I
is perefctly random, irrespective of the order in 
which the agents make measurements on the entangled qubits; also, 
each agent is equally likely to become the leader. 
In this section we attempt to mimic similar unbiased and fair election 
and we propose a Protocol II which succeeds in certain cases in a randomized
sense. To this effect, we design our Protocol II by making
the agents together play the role of an adversary to the referee; 
the leader declared by the referee may be biased if an agent so chooses.
The evil designs of the agents may however not work always and 
therefore a finite probability of fair and unbiased election may result.

What really is fairness and bias in this scenario, respecting the overall 
definition we gave earlier in Section \ref{framework}?
Firstly, fairness forces us to design protocols that treat all agents 
equally, providing them with equal opportunity in the
election process, statistically speaking. This must be ensured at the 
referee's end. Secondly, bias can be introduced in the choice presented 
by an agent so that individual agents have 
unequal probabilities of becoming elected as leader.
So, for unbiased election, we may expect agents to honestly use their
ideal random bit generating capability. A biased agent (adversary) may 
not use its perfect random bit generator and may instead bias all
his outcomes in a protocol. Providing such biased bits to the referee
would lead to a biased election where all the agents may not have equal
probability of getting elected. So, we propose our Protocol II where
each agent sends just one bit (biased or perfectly ideal random bit)
to the referee. Being honest and fair, the referee collects all 
these $n$ bits from the $n$ agents and then selects a perfectly random sample
of $\log n$ bits from the $n$ collected bits to compute the index $m$ of the
leader $A_m$. Initiation of the protocol and the agreement 
step are as in Protocol I. This completes the description of Protocol II.

We now show that biased adversary agents can not fool the referee in an 
asymptotic and statistical sense,
if there are $k=\log n$ biased agents.
We also show that a larger number of biased agents, proportional to 
$n/\log n$, can turn the table the other way; in this case the 
probability that the referee can not vanquish the evil designs of the biased 
agents is a nonzero constant. We state this as a theorem.

\begin{theorem}
It is possible to design a randomized protocol that uses no quantum 
entanglement,
but uses perfect random bit generating capability in each of the $n$ agents and
only $O(n)$ bits of classical communication so that 
as $n$ grows asymptotically, (i) fair and unbiased election of a leader is
guaranteed with probability approaching 1 if there are only $O(\log n)$ 
biased agents, and (ii) unbiased election is possible only
with a nonzero finite limiting probability if the number of biased agents is
$O(n/\log n)$. 
\label{biastheorem}
\end{theorem}

\begin{proof} 
The referee collects the single bit votes sent by the 
agents, each agent sending one bit vote. To be fair to all, the sincere
referee picks up a random sample of $\log n$ bits from the $n$ votes and 
arranges the bits in random order to create the $\log n$ bit address of 
the leader who is thus deemed elected. Note here that as long as each
agent is honestly sending a vote 
which is a perfect random bit, the address generated 
for the leader will be fair and statistically uniform over all agents.
So, as long as the agents themselves are honest, leader election is fair
even about who becomes the leader. The 
problem arises when the bit sent by a dishonest or 
biased agent is not perfectly random. If the referee happens to select any
such biased bit in his random sample for the leader's address, then the 
leader elected is statistically biased, not uniform over all agents.
We note that the probability that the random sample would be able to avoid 
biased bits is $p(n,k)=C(n-k,\log n)/C(n,\log n)$ if there are $k<n-\log n$ 
biased bits. We show in Appendix that $p(n,\log n)$ approaches 1 as $n$ grows
asymptotically. Thus, we conclude that as long as there are only a small
number of liars amongst the agents who attempt to bias leader election by 
casting biased votes, the referee succeeds in deciding upon a
leader without bias with very high probability approaching unity.
We also show in Appendix that $p(n,n/\log n)$ approaches a non-zero constant
as $n$ grows asymptotically, showing that there is a non-zero probability that 
the referee fails to deliver unbiased election when the number of liars biasing 
their votes is as large as $O(n/\log n)$. 
\end{proof}

\section{New Complexity Classes}
\label{newclass}
In the context of our results (Theorems \ref{quantumprotocoltheorem} 
and \ref{biastheorem}), 
we define two new complexity classes
where the $n$ agents are required to use at most linear classical 
communication but may use some quantum resource or randomized resource.
We say that a problem belongs the class $QECC(g, n)$ if the agents 
use $O(g)$ $ebits$ and $O(n)$ $cbits$. An $ebit$ is equivalent to 
an $n$-partite maximally entangled state shared between the agents as we have
used 
in Section \ref{quantumelection}. We saw in our 
Theorem \ref{quantumprotocoltheorem} how the fair 
and unbiased election of a leader is possible in 
$QECC(\log n, n)$. We define another class $RCC(r, n)$, 
of problems that can be solved 
using $n$ agents that share an $O(r)$-bit random string and use 
only linear classical communication complexity. It is easy to observe that 
$RCC(r,n)\subseteq QECC(r,n)$. This is because by measuring his own qubits, one qubit in each of the ebits shared with other agents, an agent can generate as
many ideal random bits shared between all the agents.  
Whether this relationship is a proper containment is an
interesting fundamental open problem concerning the
power of quantum entanglement as a computational and communication
resource.

\section{Concluding Remarks}
\label{conclusion}

It would be interesting to design other classical and randomized
protocols for fair and unbiased leader election that work for any number
of liars, possibly using superlinear communication complexity. 
Other fundamental problems in distributed computing related to multicasting,
broadcasting or consensus formation may also be studied to see if quantum 
entanglement can reduce classical communication requirement in asymptotically 
randomized protocols.

\section*{Appendix}

Let $C(n,r)$ be the number of combinations of $n$ things taken $r$ at 
a time without repetitions. We note that 

$${\frac {C(n-k,\log n)} {C(n,\log n)}}=
{\frac {(n-k)! / (\log n)! (n-k-\log n)!} {n!/(\log n)! (n-\log n)!}}$$ 
\noindent Cancelling the common terms 
and applying Sterling's approximation to the
four terms for large $n>>k$, we have the estimate
$${\frac {(n-k)^{n-k+{\frac 1 2}} (n-\log n)^{n-\log n+{\frac 1 2}} }     
 { (n-k-\log n)^{n-k-\log n+{\frac 1 2}}   n^{n+{\frac 1 2}}}  }$$
$$={\frac     { n^{n-k+{\frac 1 2}}
 (1-{\frac k n})^{n-k+{\frac 1 2}} 
(n-\log n)^{n-\log n+{\frac 1 2}}         }       
{(n-\log n)^{n-k-\log n+{\frac 1 2}} 
(1-{\frac k {n-\log n}})^{n-k-\log n+{\frac 1 2}}    
n^{n+{\frac 1 2}} }   }$$
$$={\frac {
  (1-{\frac {\log n} n})^k
 (1-{\frac k n})^{n-k+{\frac 1 2}}
          }
          {
(1-{\frac k {n-\log n}})^{n-k-\log n+{\frac 1 2}}    
          } 
    }   $$

Now note that putting $k=\log n$, and using Euler's formula with $n$
increasing asymptotically, the above estimate approaches unity. On the 
other hand, putting $k={\frac {Cn} {\log n}}$, for 
a constant $C>0$ takes the estimate 
to $e^{-C}$ as $n$ blows up. Here $e$ is Euler's constant.

\end{document}